\documentclass[aps,twocolumn,showpacs]{revtex4}
\usepackage{graphicx}

\begin{document}

\title{Observation of dynamical instability for a Bose-Einstein\\
condensate in a moving 1D optical lattice}

\author{L. Fallani$^{*}$, L. De Sarlo, J. E. Lye, M. Modugno$^{1}$,
R. Saers$^{\dagger}$, C. Fort, and M. Inguscio}

\affiliation{LENS, Dipartimento di Fisica, and INFM Universit\`a
di Firenze via Nello
Carrara 1, I-50019 Sesto Fiorentino (FI), Italy\\
$^{1}$ also BEC-INFM Center, Universit\`a di Trento, I-38050 Povo
(TN), Italy}

\begin{abstract}
We have experimentally studied the unstable dynamics of a
harmonically trapped Bose-Einstein condensate loaded into a 1D
moving optical lattice. The lifetime of the condensate in such a
potential exhibits a dramatic dependence on the quasimomentum
state. This is unambiguously attributed to the onset of dynamical
instability, after a comparison with the predictions of the
Gross-Pitaevskii theory. Deeply in the unstable region we observe
the rapid appearance of complex structures in the atomic density
profile, as a consequence of the condensate phase uniformity
breakdown.
\end{abstract}

\pacs{03.75.Kk, 03.75.Lm, 32.80.Pj, 05.45.-a}

%\date{\today}

\maketitle

Ultracold atoms in optical lattices have proven to be a rich field
of investigation for both fundamental and applicative issues of
quantum mechanics. Periodic potentials are very well known in
solid state physics, where the basic description of the system is
the Bloch theory for a gas of noninteracting particles, such as
electrons in crystalline solids \cite{solidstate}. Cold atoms in
optical lattices have opened the possibility to investigate
effects not previously observable on ordinary matter crystals,
such as Bloch oscillations, Wannier-Stark ladders and Landau-Zener
tunnelling \cite{blochosc}. Bose-Einstein condensates (BECs) are
particularly well suited for the investigation of these phenomena,
due to the small momentum spread and the large coherence length.
However, in a high density sample, such as a trapped BEC,
nonlinearities induced by interactions among the atoms can
complicate the picture and lead to a number of new effects, such
as the observation of new quantum phases \cite{greiner}, the
generation of solitonic propagation \cite{eiermann} and the
observation of different kinds of instabilities. In the recent
years many papers have investigated the latter topic, both
theoretically \cite{wu,smerzi,machholm,nesi,konotop,dalfovo} and
experimentally \cite{burger,cataliottinjp,pisani}. Many of these
works have suggested that the BEC superflow can be broken not only
by the mechanism of energetic (Landau) instability, but also by
dynamical instability, occurring for condensates with repulsive
interactions in periodic potentials.

In this Letter we report on the unambiguous observation of
dynamical instability for a trapped BEC in a moving 1D optical
lattice. We have measured the characteristic rates for this
instability to occur, mapping the quasimomentum space for both the
lowest and excited energy bands. This quantitative analysis gives
important information on the timescales limiting the investigation
of coherence effects in a BEC moving in a periodic potential.

Dynamical instability is a peculiar feature of nonlinear systems.
It occurs when the eigenspectrum of the excitations of the system
exhibits complex frequencies. In this case, arbitrary small
perturbations of the wavefunction may grow exponentially,
eventually leading to the destruction of the initial state. The
conditions for this kind of instability are satisfied by a BEC
with repulsive interactions in a periodic potential. When the
height of the periodic potential is not too large, the superfluid
ground state of a BEC in an optical lattice is well described by
the Gross-Pitaevskii equation (GPE)
\begin{eqnarray}
i\hbar \frac{\partial \Psi}{\partial t} & = & \left(
-\frac{\hbar^2}{2m}\nabla^2 + V_{trap}(\vec r)
\right. \nonumber \\
& & \left. + sE_R \cos ^2(kx) + \frac{4\pi \hbar^2 a}{m}|\Psi|^2
\right) \Psi, \label {GPE}
\end{eqnarray}
where $\Psi$ is the complex BEC order parameter, $V_{trap}(\vec
r)$ is the harmonic trapping potential, $s$ is the height of the
periodic potential in units of the recoil energy $E_R=\hbar^2
k^2/2m$, $\pi /k$ is the periodicity of the lattice in real space,
and $a$ is the scattering length. Among the stationary solutions
of Eq.~(\ref{GPE}) there are the usual Bloch waves, i.e. plane
waves with quasimomentum $q$ and band index $n$, modulated in
space by functions having the same periodicity of the lattice. A
linear stability analysis of the GPE shows that, in certain
regions of the quasimomentum space, these solutions are not
dynamically stable and an exponential growth of perturbations may
start \cite{wu,smerzi,dalfovo}. The onset of dynamical instability
has been considered as a possible explanation for the disruption
of the atomic superflow in a past experiment \cite{cataliottinjp},
concerning large amplitude dipole oscillations in the magnetic
trap + static optical lattice. In this work we quantitatively
investigate the unstable dynamics of a trapped BEC in a moving
optical lattice in the regime of low lattice heights. The novelty
of this work, with respect to the previous experiments
\cite{burger,cataliottinjp}, is that the implementation of a
lattice moving at constant velocity allows us to accurately set
the quasimomentum of the condensate and to make a precise
investigation of the stability regimes in the full Brillouin zone
and for different energy bands.

\begin{figure}[t!]
\begin{center}
\includegraphics[width=\columnwidth]{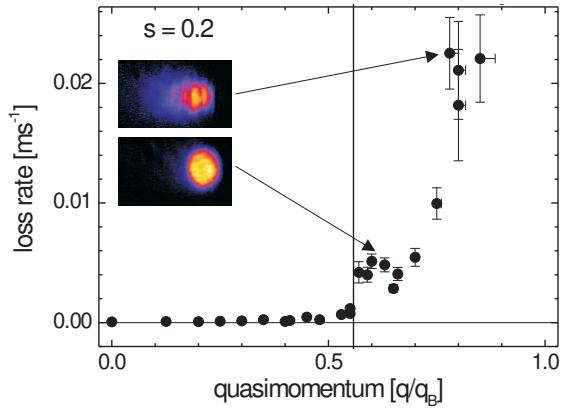}
\end{center}
\caption{Loss rates for a trapped BEC loaded in a moving optical
lattice with $s=0.2$. The vertical line corresponds to the
calculated threshold for the onset of dynamical instability
\cite{dalfovo}. The images show the density distribution of the
expanded cloud. Near the zone boundary, where instability is
faster, we observe the appearance of some complex structures,
evidencing the loss of coherence in the BEC.} \label{rateszoom}
\end{figure}

The experiment is performed with an elongated $^{87}$Rb
Bose-Einstein condensate produced in a Ioffe-Pritchard magnetic
trap by means of forced RF evaporation. The trapping frequencies
are $\omega_z=2\pi \times 8.8$ Hz axially and $\omega_\perp=2\pi
\times 91$ Hz radially, with the axis of the trap oriented
horizontally. Our typical condensates are made of $\simeq 3 \times
10^5$ atoms in the hyperfine ground state $|F=1;m_F=-1>$, with a
peak density $n\simeq1.2 \times 10^{14}$ cm$^{-3}$. The optical
lattice is created by the interference of two counterpropagating
laser beams derived from the same Ti:Sa laser operating at
$\lambda=820$ nm, far detuned with respect to the Rb D1 line at
$\lambda=795$ nm. The lattice beams are aligned along the symmetry
axis of the condensate and are only slightly focused (400 $\mu$m
diameter), so that the optical radial confinement may be
completely neglected. The frequencies of the two beams are
controlled by two acousto-optic modulators (AOMs) driven by two
phase-locked radiofrequency generators in order to provide a
stable detuning $\delta\nu$ between the two beams. The resulting
interference pattern, averaged over the optical frequencies, is a
standing wave moving at velocity $v=(\lambda/2) \delta\nu$. Once
the condensate has been produced, we adiabatically switch on the
moving lattice loading the condensate in a state of well defined
quasimomentum $q=mv/\hbar$ and band index $n$ \cite{lensing}. We
let the BEC evolve in this potential for a variable time, then we
switch off both the magnetic trap and the optical lattice and,
after an expansion of 28 ms, we image the atomic cloud along the
radial horizontal direction \cite{nota2}.

The number of atoms remaining in the condensate decreases
exponentially as a function of the time spent in the periodic
potential. We have measured the lifetime of the condensate for
different values of the quasimomentum $q$ and for different energy
bands. Since even a tiny thermal component may seriously limit the
BEC lifetime, we have used an RF-shield in order to remove the
hottest atoms produced by heating of the atomic sample. In this
way we measure lifetimes of the order of $\approx 10$ s in the
lattice at $v=0$, with no discernable thermal fraction even on
long timescales.

\begin{figure}[b!]
\begin{center}
\includegraphics[width=\columnwidth]{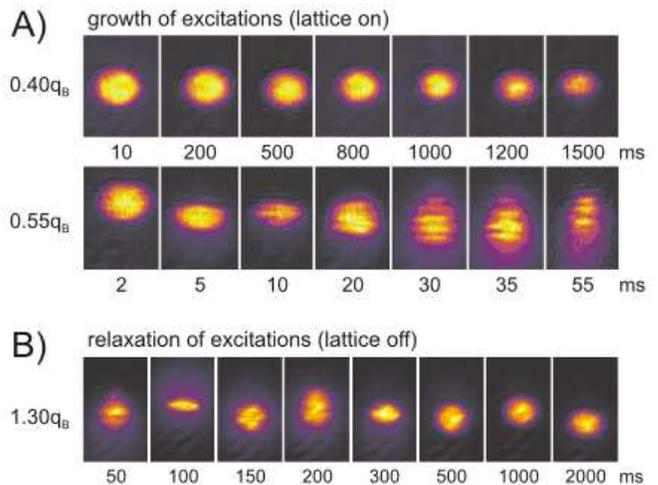}
\end{center}
\caption{A) Evolution of the expanded atomic density profile for a
lattice height $s=1.15$ and two different quasimomentum values
$q=0.4q_B$ and $q=0.55q_B$ below and above the threshold for
dynamical instability. Note the different timescales. B) Evolution
of the expanded atomic density profile for a variable time spent
in the pure magnetic trap, following 5 ms spent in an optical
lattice with $s=1.15$ and $q=1.30q_B$. In all these pictures the
lattice is directed from top to bottom.} \label{stripes}
\end{figure}

The experimental results are analyzed according to the GPE shown
in Eq.~(1). As a matter of fact, one can avoid the complication of
the full 3D theory by using the Non Polynomial Schr\"odinger
equation (NPSE), a simplified 1D model that includes an effective
radial-to-axial coupling \cite{salasnich}.  The NPSE provides a
more realistic description of the actual system with respect to
the simple one dimensional GPE used previously \cite{wu}, yielding
an estimate of the instability thresholds and of the growth rates
of the most unstable modes in nice agreement with that of Eq.~(1)
in the regime of lattice heights considered here (see
\cite{dalfovo} for details).

In Fig.~\ref{rateszoom} we show the measured loss rates (inverse
of the lifetime) for the BEC in an optical lattice with $s=0.2$ as
a function of $q$. With increasing $q$, from the bottom of the
first band to the zone boundary, the lifetime changes
dramatically, spanning three orders of magnitude from $\approx 10$
s to $\approx 10$ ms. In particular, we observe a discontinuity
around $q=0.55q_B$ (where $q_B=2 \pi/\lambda$ is the boundary of
the first Brillouin zone), in good correspondence with the
calculated threshold for the onset of dynamical instability
(vertical line). Deeply in the dynamically unstable regime, as
shown in the images of Fig.~\ref{rateszoom}, we observe the
appearance of some complex structures in the expanded BEC density
profile, suggesting fragmentation of the Bloch wave. These
interference-like structures are particularly visible for higher
lattice heights and near the zone boundaries, where the evolution
of instability is faster. In Fig.~\ref{stripes}A we present two
sequences of images showing the time evolution of the atomic
density profile for a lattice height $s=1.15$ and two different
values of quasimomentum below and above the threshold for
dynamical instability. This behavior cannot be attributed to a
simple heating of the sample. On the contrary, it may reflect the
creation of phase domains induced by the growth of instabilities,
which break the phase uniformity of the BEC \cite{nesi}. When
these structures are not just weak perturbations of the density
profile, it is difficult to measure the number of atoms in the
condensate directly. In order to measure it more precisely, we let
the cloud evolve in the pure magnetic trap, allowing relaxation of
the excitations. We have measured typical relaxation times of the
order of $\approx 500$ ms, after which the BEC recovers its smooth
density profile, as shown in Fig.~\ref{stripes}B. This observation
suggests the existence of mechanisms that, once the cause of
instability is removed, allow the system to come back to the
ground state, damping the excitations and restoring the initial
coherence.

\begin{figure}[t!]
\begin{center}
\includegraphics[width=0.85\columnwidth]{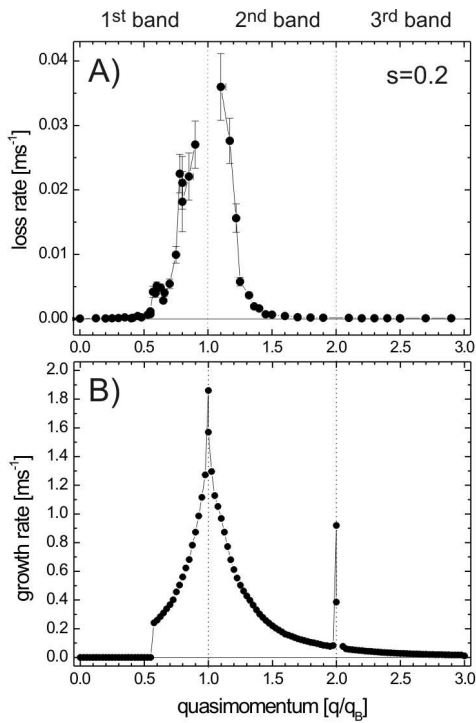}
\end{center}
\caption{A) Experimental loss rates for a BEC loaded into the
first three energy bands of a moving optical lattice with $s=0.2$.
B) Theoretical growth rates of the most dynamically unstable modes
obtained from a linear stability analysis of the NPSE
\cite{dalfovo}.} \label{rates02}
\end{figure}

\begin{figure}[t!]
\begin{center}
\includegraphics[width=0.85\columnwidth]{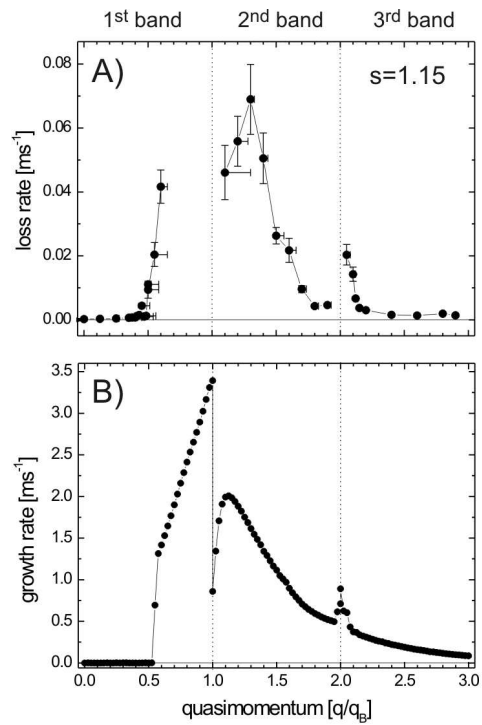}
\end{center}
\caption{A) Experimental loss rates for a BEC loaded into the
first three energy bands of a moving optical lattice with
$s=1.15$. B) Theoretical growth rates of the most dynamically
unstable modes obtained from a linear stability analysis of the
NPSE \cite{dalfovo}.} \label{rates115}
\end{figure}

Our experimental technique allows us to study the dynamics in the
higher energy bands and compare the peculiar behavior in each band
with the results of the theoretical model. The remarkable
agreement between the theory and the experiment enforces that the
observed phenomena are indeed due to dynamical instability. In
Fig.~\ref{rates02}A we show the experimental loss rates measured
in the lowest three energy bands as a function of the
quasimomentum $q$ for $s=0.2$. These rates are compared in
Fig.~\ref{rates02}B with the theoretical growth rates of the most
dynamically unstable modes, obtained from the linear stability
analysis of the NPSE \cite{dalfovo}. Increasing the height of the
optical lattice from $s=0.2$ to $s=1.15$, as shown in
Fig.~\ref{rates115}, the picture in the higher bands starts to
develop asymmetric features around the zone edges. This is
particularly evident crossing the boundary between the second and
the third Brillouin zone, where we observe a marked change in the
instability rates. This nontrivial behavior cannot be attributed
to single particle band structure considerations (for example,
residual nonadiabaticites or interband transitions). Indeed,
comparing this feature with the theoretical calculations for the
growth rates of the most unstable modes, we do observe the same
distinctive shape. The experimental loss rates should not be
quantitatively compared with the theoretical growth rates, since
these two quantities have different physical meanings: the first
measures how fast the atoms are removed from the condensate; the
second is the rate at which the most unstable mode grows in the
linear regime, i.e. at the onset of dynamical instability. Out of
the linear regime, when excitations have grown and the unstable
modes are not just weak perturbations of the carrier Bloch wave,
the dynamical evolution of the system cannot be explained with
this perturbative approach and the full solution of the
time-dependent 3D-GPE is required. However, the remarkable
similarity between the experimental and theoretical curves
indicates that the onset of the instability produces a significant
imprinting on the subsequent dynamics of the system.

In the experiment, the BEC is trapped in the harmonic magnetic
potential, in order to have a high density sample and long
observation times. On the other hand, the presence of the harmonic
potential may induce a variation of the quasimomentum in time,
thus changing the conditions for instability. In a semiclassical
approach, a force $F$ acting on the system results in a variation
of the quasimomentum $q$ according to the law $F=dq/dt$.
Integrating this equation of motion with the velocity spectrum
$v(q)$ of the Bloch bands, one can demonstrate that the
quasimomentum $q$ makes small oscillations at the axial trap
frequency (rescaled with the effective mass) \cite{nota1}. We have
taken this effect into account by including horizontal error bars
in Figs.~\ref{rates02} and \ref{rates115}, representing the
maximum variation of $q$ during the experiment. In real space
these oscillations correspond to a micro-motion of the BEC around
the center of the trap with an amplitude much less than 1 $\mu$m.
Actually, this effect becomes important only when the Bloch bands
significantly differ from the free particle energy, i.e. near the
zone boundaries and for higher lattice heights.

Repeating the experiment in the presence of a small thermal
component we observed that, even for small values of quasimomentum
($q \gtrsim 0.05 q_B$), the atomic sample is completely destroyed
in a time much shorter than the lifetime measured for the pure
BEC. We attribute this behavior to the onset of energetic
instability, that occurs in the presence of dissipative processes
\cite{wu,dalfovo}. The residual thermal fraction surrounding the
BEC can provide a mechanism for such a dissipation, thus
triggering the activation of energetic instability. The study of
this regime will be the subject of further experimental work.

In conclusion, we report the observation of dynamical instability
for a harmonically trapped Bose-Einstein condensate in the
presence of a 1D optical lattice. The ability to precisely control
and change the lattice velocity has allowed us to carry out a
detailed investigation of the instability regimes in the full
Brillouin zone for different energy bands. We have quantitatively
studied the lifetime of the condensate in such a potential,
finding a good agreement between the experimental loss rates and
the theoretical rates for the growth of excitations at the onset
of dynamical instability. Deeply in the dynamically unstable
regime we have observed the appearance of complex structures in
the expanded density profile, a signature of a loss of coherence
in the atomic sample. These observations clearly identify
dynamical instability as the main mechanism responsible for the
disruption of the matter wave superflow in the regime of small
lattice heights. The quantitative analysis presented in this
letter gives important information on the characteristic
timescales which limit the investigation of the coherence
properties of a BEC moving in a periodic potential.

This work has been supported by the EU under Contracts No. HPRI-CT
1999-00111 and HPRN-CT-2000-00125 and by the INFM Progetto di
Ricerca Avanzata ``Photon Matter''. J.~E.~L. was supported by EU
with a Marie Curie Intra-European Fellowship. We thank F. S.
Cataliotti for stimulating discussions.


\begin{thebibliography}{99}

\bibitem[*]{mail}
Electronic address: \verb"fallani@lens.unifi.it"

\bibitem[$\dagger$]{robert}
Now at Department of Physics, Ume\aa~University, S-901 87 Ume\aa,
Sweden.

\bibitem{solidstate}
F.~Bloch, Z. Phys. \textbf{52}, 555 (1929); C.~Zener, Proc. R.
Soc. London A \textbf{145}, 523 (1934).

\bibitem{blochosc}
M.~Raizen, C.~Salomon and Q.~Niu, Phys. Today \textbf{50}, 30
(1997) and references therein.

\bibitem{greiner}
M.~Greiner, O.~Mandel, T.~Esslinger, T.~W.~H\"{a}nsch, I.~Bloch,
Nature \textbf{415}, 39 (2002).

\bibitem{eiermann}
B.~Eiermann, Th.~Anker, M.~Albiez, M.~Taglieber, P.~Treutlein,
K.-P.~Marzlin, and M.~K.~Oberthaler, eprint cond-mat/0402178
(2004).

\bibitem{wu}
B.~Wu and Q.~Niu, Phys. Rev. A \textbf{64}, 061603R (2001); B.~Wu
and Q.~Niu, New Journ. Phys. \textbf{5}, 104 (2003).

\bibitem{nesi}
F.~Nesi and M.~Modugno, J. Phys. B \textbf{37}, S101 (2004).

\bibitem{smerzi}
A.~Smerzi, A.~Trombettoni, P.~G.~Kevrekidis and A.~R.~Bishop,
Phys. Rev. Lett. \textbf{89}, 170402 (2002); C.~Menotti, A.~Smerzi
and A.~Trombettoni, New Journ. Phys. \textbf{5}, 112 (2003).

\bibitem{machholm}
M.~Machholm, C.~J.~Pethick and H.~Smith, Phys. Rev. A \textbf{67},
053613 (2003).

\bibitem{konotop}
F.~Kh.~Abdullaev, B.~B.~Baizakov, S.~A.~Darmanyan, V.~V.~Konotop
and M.~Salerno, Phys. Rev. A \textbf{64}, 043606 (2001);
V.~V.~Konotop and M.~Salerno, Phys. Rev. A \textbf{65}, 021602R
(2002).

\bibitem{dalfovo}
M.~Modugno, C.~Tozzo and F.~Dalfovo, in preparation (2004).

\bibitem{burger}
S.~Burger, F.~S.~Cataliotti, C.~Fort, F.~Minardi, M.~Inguscio,
M.~L.~Chiofalo, M.~P.~Tosi, Phys. Rev. Lett. \textbf{86}, 4447
(2001); B.~Wu and Q.~Niu, Phys. Rev. Lett. \textbf{89}, 088901
(2002); S.~Burger, F.~S.~Cataliotti, C.~Fort, F.~Minardi,
M.~Inguscio, M.~L.~Chiofalo, M.~P.~Tosi, Phys. Rev. Lett.
\textbf{89}, 088902 (2002).

\bibitem{cataliottinjp}
F.~S.~Cataliotti, L.~Fallani, F.~Ferlaino, C.~Fort, P.~Maddaloni,
M.~Inguscio, New Journ. Phys. \textbf{5}, 71 (2003).

\bibitem{pisani}
M.~Cristiani, O.~Morsch, N.~Malossi, M.~Jona-Lasinio,
M.~Anderlini, E.~Courtade and E.~Arimondo, Optics Express
\textbf{12}, 4 (2004).

\bibitem{lensing}
L.~Fallani, F.~S.~Cataliotti, J.~Catani, C.~Fort, M.~Modugno,
M.~Zawada and M.~Inguscio, Phys. Rev. Lett. \textbf{91}, 240405
(2003).

\bibitem{nota2}
We note that in all the experiments the lattice switching off is
slow enough to ensure adiabaticity.

\bibitem{salasnich}
L.~Salasnich, Laser Physics \textbf{12}, 198 (2002); L.~Salasnich,
A.~Parola, and L.~Reatto, Phys. Rev. A \textbf{65}, 043614 (2002).

\bibitem{nota1}
Actually, for initial quasimomentum states near the region in
which the effective mass diverges, the system may enter a Bloch
oscillations-like regime, in which the quasimomentum $q$, instead
of oscillating in time, grows indefinitely. The experimental
points plotted in the graphs do not refer to this regime.

\end{thebibliography}
\end{document}